\documentclass[aps,prl,amsmath]{revtex4}
\usepackage{graphicx}
\usepackage{color}
\usepackage{bm}
\usepackage{epsfig}
\usepackage{latexsym}
\usepackage{pifont}
\usepackage{float}
\usepackage[utf8]{inputenc}
\graphicspath{{figure/}}
\usepackage{siunitx}
\usepackage{textgreek}

\begin{document}
\newcommand{\be}{\begin{equation}}
\newcommand{\ee}{\end{equation}}
\newcommand{\bea}{\begin{eqnarray}}
\newcommand{\eea}{\end{eqnarray}}
\newcommand{\nn}{\nonumber}
\renewcommand{\figurename}{Fig.}

\title{Shape dynamics and migration of branched cells on complex networks}
\author{Jiayi Liu$^{1,2}$, Javier Boix-Campos$^{3}$, Jonathan E. Ron$^{1,4}$, Johan M. Kux$^{3}$, Nir S. Gov$^{1}$ and Pablo J. S\'{a}ez$^{3}$}
\affiliation{$^{1}$Department of Chemical and Biological Physics, Weizmann Institute of Science, Rehovot, Israel}
\affiliation{$^{2}$Department of Physics, Yale University, New Haven, CT, USA}
\affiliation{$^{3}$Cell Communication and Migration Laboratory, Institute of Biochemistry and Molecular Cell Biology, Center for Experimental Medicine, University Medical Center Hamburg-Eppendorf, Hamburg, Germany}
\affiliation{$^{4}$Department of Physics, Technion, Haifa, Israel}

\begin{abstract}
Migratory and tissue resident cells exhibit highly branched morphologies to perform their function and to adapt to the microenvironment. Immune cells, for example, display transient branched shapes while exploring the surrounding tissues. In another example, to properly irrigate the tissues, blood vessels bifurcate thereby forcing the branching of cells moving on top or within the vessels. In both cases microenvironmental constraints force migrating cells to extend several highly dynamic protrusions. Here, we present a theoretical model for the shape dynamics and migration of cells that simultaneously span several junctions, which we validated by using micropatterns with an hexagonal array, and a neuronal network image analysis pipeline to monitor the macrophages and endothelial cell shapes and migration. In our model we describe how the actin retrograde flow controls branch extension, retraction and global cell polarization. We relate the noise in this flow to the residency times and trapping of the cell at the junctions of the network. In addition, we found that macrophages and endothelial cells display very different migration regimes on the network, with macrophages moving faster and having larger changes in cell length in comparison to endothelial cells. These results expose how cellular shapes and migration are intricately coupled inside complex geometries.  
\end{abstract}

\maketitle

\section*{Introduction}

Cells migrate within tissues during physiological and pathological processes, such as  embryo development, tissue growth/repair, immune response, and cancer propagation \cite{paul2017cancer,yamada2019mechanisms,kameritsch2020principles,pawluchin2022moving}. Due to the constrains of the microenvironment, motile cells often need to navigate tissues and organs avoiding the surrounding obstacles \cite{paksa2016repulsive}. In order to efficiently migrate, immune \cite{georgantzoglou2022two,peterman2023zebrafish} and cancer \cite{farin2006transplanted,paul2017cancer} cells acquire branched morphologies that allow them to explore the microenvironment. Cell migration in complex geometries has been studied in different \textit{in vitro} models using different geometries and types of confinement \cite{ambravaneswaran2010directional,pham2018cell,um2019immature,renkawitz2019nuclear,zhao2019cell,belotti2020analysis,wang2020chemotaxing,monzo2021adaptive,hadjitheodorou2022mechanical,ron2024emergent}. Once migrating cells encounter junctions they generate protrusions along alternate directions \cite{dai2020tissue}, and eventually select a new direction \cite{kameritsch2020principles,ron2024emergent}. How cells efficiently migrate through complex geometries despite the large number of branches is still poorly understood.

We have recently shown how cellular shape and cytoskeleton dynamics control the migrating behavior of cells facing a single symmetric Y-junction \cite{ron2024emergent}. Here, we generalize our previous theoretical model \cite{ron2024emergent} to describe cells that are moving on a hexagonal network where the cells can simultaneously span multiple junctions (Fig. \ref{fig1}A-F). Inside the body similar geometries are encountered by endothelial cells and macrophages moving over and inside blood vessels \cite{fantin2010tissue}, during angiogenesis \cite{fonseca2020endothelial} and during the immune response \cite{barros2017live}. 
Using live-cell imaging of human umbilical vein endothelial cells (HUVEC), bone marrow-derived macrophages, and hexagonal micropatterns, we monitored the migration patterns and cellular shape dynamics of cells that are spread over multiple junctions (Fig. \ref{fig1}G), to compare with the theoretical model.

\section*{Model}

\begin{figure}[b!]
    \centering
    \includegraphics[width=1\textwidth]{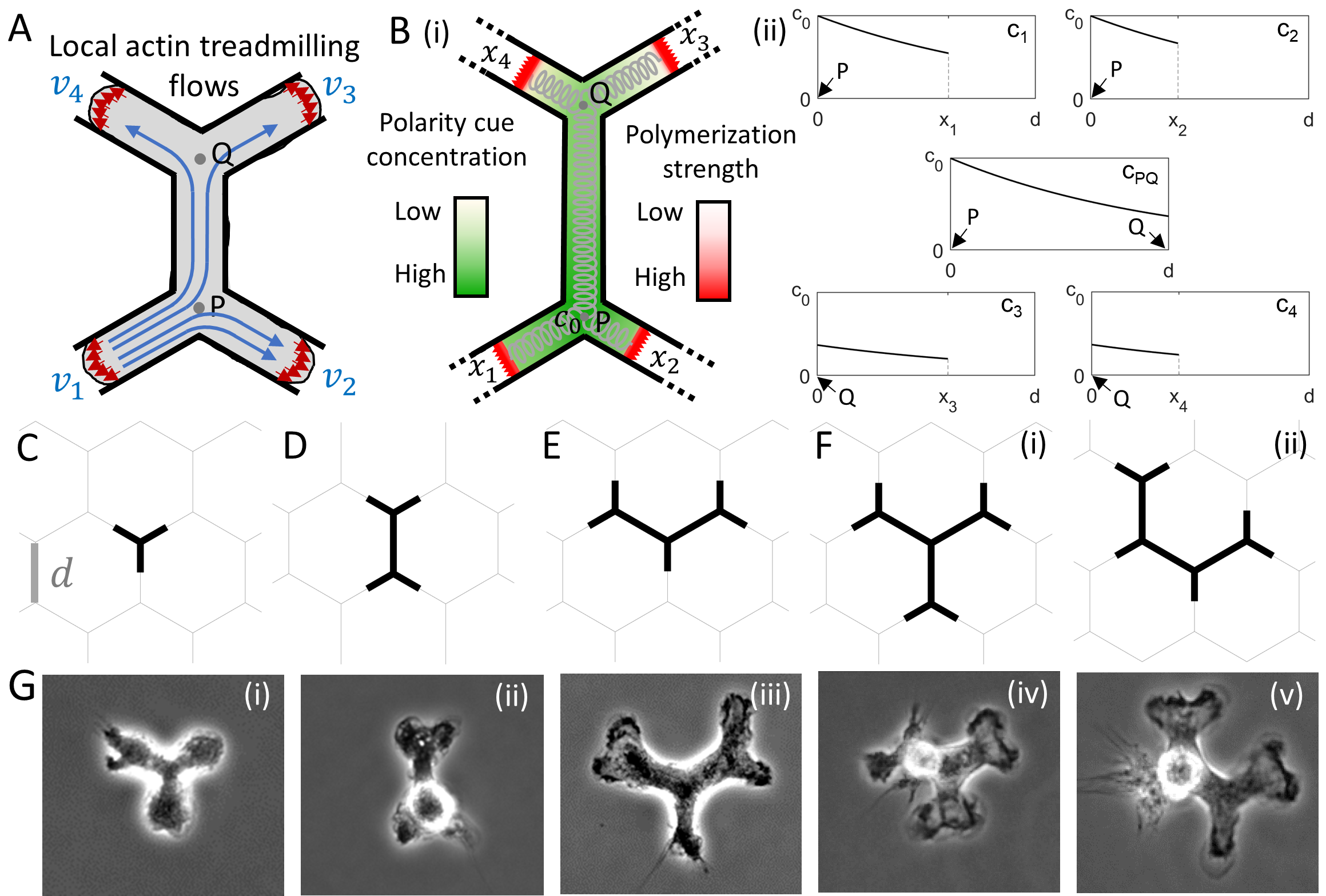}
    \caption{ Multiple junction model. (A) Scheme showing the local actin treadmilling flows at each edge of the arms, and the split of the local actin flow that emanates from arm 1: it undergoes symmetric splits at each junction. (B) (i) An example of the concentration field of the polarity cue, which is affected by the advection flows. The cell elasticity is denoted by the grey springs. (ii) Polarity cue concentration profiles of all the segments in (i), where $d$ is the length of the hexagonal network segments. Node P is the reference point where the polarity cue concentration is fixed to be \(c_0\) (Eq.\ref{c_i_2junc}), and node Q is the other junction node that the cell is spanning. (C-F) Shape of the cells that are spanning different number of junctions. (C) One junction. (D) Two junctions. (E) Three junctions. (F) Four Junctions, where (i) and (ii) are the two possible shapes of the cell. G) Snapshots of macrophages spanning multiple junctions. (i)-(v) correspond to the shapes C, D, E, F(i) and F(ii) respectively. Hexagon side in snapshots (i-iii): 12.5~\textmu m. Hexagon side in snapshots (iv-v): 20~\textmu m. Inter-hexagon width in snapshots (i-iii): 20~\textmu m. Inter-hexagon width in snapshots (iv-v): 24~\textmu m. }
    \label{fig1}
\end{figure}

Our model is based on a cell composed of one-dimensional sections, which spontaneously self-polarizes due to the coupling between the local actin polymerization activity at the leading edges of the cellular protrusions (Fig. \ref{fig1}A) \cite{ron2020one,ron2024emergent}. This coupling is mediated by the advection of an inhibitor of actin polymerization (termed "polarity cue") by the net actin retrograde flow (Fig. \ref{fig1}B). The cell polarizes when the frontal edge of the cell has the largest actin polymerization activity, which is maintained by the actin flow that advects the inhibitor to the remaining protrusions, where actin polymerization is inhibited. 

We extend here our previous model for cells moving over a symmetric Y-shaped junction \cite{ron2024emergent} to describe cells moving on a symmetric hexagonal network of junctions where the cells can span up to four junctions simultaneously (Fig. \ref{fig1}C-F). The model could also be extended to describe cells spanning a larger number of junctions.

The polymerization of actin at the cellular tips is converted into protrusive forces, which compete with elastic and friction forces. This force balance drives the extension/retraction of the cellular protrusions and cause the cell to migrate. For simplicity, we assume that all the forces which act on each cellular protrusion are localized at their tips (frontal edges). The global elasticity of the cell is treated as an elastic spring.

For a cell with \(N\) \((N \geq 3)\) arms, i.e., spanning across \(N-2\) junctions (Fig. \ref{fig1}C-F), the dynamics of arm \(i\) is described by three variables \cite{ron2024emergent}: 1) its length \(x_i\), 2) the fraction of active slip-bonds adhesion \(n_i\) at the leading edge, and 3) the local actin treadmilling flow velocity \(v_i\) at the leading edge (Fig. \ref{fig1}E). The dynamic equations for these variables are given by

\begin{equation} \label{x_i}
{\dot{x}}_i = \frac{1}{\Gamma_i} [v_i - k \, (L-1)]
\end{equation}

\begin{equation} \label{n_i}
{\dot{n}}_i = r \, (1-n_i) - n_i \, \exp{\left[\frac{-v_i+k\,(L-1)}{f_s\,n_i}\right]}
\end{equation}

\begin{equation} \label{v_i}
{\dot{v}}_i = -\delta \, (v_i-v_i^*) +  \sigma \xi_t
\end{equation}

In Eq.\ref{x_i}, \(\Gamma_i\) is a non-constant friction coefficient that depends on the direction of motion of the arm's leading edge, given by

\begin{equation}
\Gamma_i = \Theta(\dot{x}_i )+ (1-\Theta(\dot{x}_i)) \, n_i \, \exp{\left[\frac{v_i-k\,(L-1)}{f_s\,n_i}\right]}
\end{equation}

where \(\Theta\) is a Heaviside function and \(\kappa\) is the effective spring constant of the bond-linkers. When the arm is extending, the friction acts as a constant drag \(\Gamma_i=1\), while when the arm is retracting, the friction is due to the adhesion of the slip-bonds. 

Eq.\ref{x_i} is a simplified description of the protrusive traction forces, which can be further elaborated to include the adhesion dependence of these forces \cite{mukherjee2022actin}. The restoring force of the cell elasticity in Eqs. \ref{x_i},\ref{n_i} is described by a simple spring term (Fig. \ref{fig1}F(i)), where \(k\) is the effective elasticity of the cell (of rest length 1). 

In Eq.\ref{n_i}, \(f_s\) describes the susceptibility of the slip-bonds to detach due to the applied force, and \(r\) is the effective cell-substrate adhesiveness. 

In Eq.\ref{v_i}, \(\delta\) is the rate at which the local actin flows relax to the steady-state solutions \(v_i^*\), which is given by \cite{ron2020one} 
\begin{equation} \label{v_ss}
v_i^* = \beta \frac{c_s}{c_s+c_i(x_i)}
\end{equation}
where $c_i(x_i)$ is the concentration of actin polymerization inhibitor at the tip of arm $i$ (at the coordinate $x_i$), $c_s$ is the saturation concentration, and \(\beta\) is the maximal actin polymerization speed at the arm edges. The spatial distribution of the inhibitor concentration is calculated below.
The term \(\sigma \xi_t\) in Eq. \ref{v_i} describes the noise in the actin polymerization activity, with a random Gaussian form of amplitude \(\sigma\).

The total length of the cell is given by 
\begin{equation}
L=\sum_{i} x_i+(N-3)\,d 
\end{equation}

where \(d\) is the distance between two adjacent junctions on the network.

\subsection*{Calculation of the actin treadmilling flows}
We assume that the local actin treadmilling flow emanating from the free edge of an arm splits symmetrically at each junction. For example, for a cell spanning 2 junctions, the flow from arm $1$, \(v_1\), splits symmetrically into arm 2 and the segment between node P and node Q (Fig. \ref{fig1}A). Considering \(v_i\) of all the arms, the net actin treadmilling flow \(u_i\) within each arm is given by the sum of the flows that enter that arm

\begin{equation} \label{u_i_2junc}
\begin{split}
u_1=v_1-\frac{v_2}{2}-\frac{v_3}{4}-\frac{v_4}{4} \\
u_2=v_2-\frac{v_1}{2}-\frac{v_3}{4}-\frac{v_4}{4} \\
u_3=v_3-\frac{v_4}{2}-\frac{v_1}{4}-\frac{v_2}{4} \\
u_4=v_4-\frac{v_3}{2}-\frac{v_1}{4}-\frac{v_2}{4}
\end{split}
\end{equation}
with the convention that a positive flow is away from the arm tip. The middle segment that connects the two adjacent junctions P and Q has a net flow from all the incoming segments that connect to it

\begin{equation}
u_m=\frac{v_3}{2}+\frac{v_4}{2}-\frac{v_1}{2}-\frac{v_2}{2}
\end{equation}
and we take it to be positive towards junction P. Generally, the net actin flow of a free arm \(i\) is given by

\begin{equation} \label{u_i}
u_i = v_i - \sum_{j \neq i} \frac{v_j}{2^{m_{i, j}}}
\end{equation}

where \(m_{i, j}\) is the number of the junction nodes between arm \(i\) and arm \(j\). Note that the relation \( \sum_{j \neq i} \frac{1}{2^{m_{i, j}}} =1 \) always holds. 

The net actin flow of a node connection segment \(k\) that connects two adjacent junction nodes Q and P is given by

\begin{equation} \label{u_k}
u_k = \sum_{i\;connect\;to\;Q} \frac{v_i}{2} - \sum_{j\;connect\;to\;P} \frac{v_j}{2}
\end{equation}
where \(\{i\}\) and \(\{j\}\) denote the free arms that emanate from junction Q and P, respectively. The positive flow of \(u_k\) is taken to be along the direction from Q to P. 

Note that within the cell, due to the nucleus and other organelles, the actin does not freely flow along the entire length of the cell, as suggested by our model (Fig. \ref{fig1}A). However, these flows drive the formation of an inhomogeneous concentration profile of the cytoplasmic contents (proteins as well as larger organelles) along the cell length, and this is captured by our simplified model.

\subsection*{Concentration profile of the polarity cue (actin polymerization inhibitor)}
We consider a polarity cue \(c(x)\) diffusing in the cytoplasm, which is advected by the net actin treadmilling flows and acts as the inhibitor of actin polymerization. We assume that the concentration profile of the polarity cue relaxes faster than the timescale of the changes in the actin flow or cell shape \cite{maiuri2015actin,lavi2016deterministic,ron2020one}. This assumption greatly simplifies the problem and allows us to use the steady-state distribution for this concentration field, which is updated as the actin flows and cell shape changes. In the steady state, the advection-diffusion equation of the polarity cue within each arm is

\begin{equation}
\frac{\partial}{\partial x} (u_i \, c_i(x) + D \, \frac{\partial c_i(x)}{\partial x}) = 0
\end{equation}
where \(u_i\) is the net actin flow in arm \(i\), \(c_i(x)\) is the concentration of the polarity cue and \(D\) is the diffusion coefficient of the polarity cue. This gives a solution
\begin{equation}
c_i(x) = c_{i,0} \, \exp{\left(-\frac{u_i x}{D}\right)} + c_{i,1}
\label{cx}
\end{equation}
By applying a no-flux boundary condition at the ends of the arms, i.e., \(u_i c_i(x)+D\frac{\partial c_i(x)}{\partial x} |_{x=x_i} =0\), we obtain the coefficient \(c_{i,1}=0\). The relationship between the coefficients \(c_{i,0}\) \((i=1,2,...,N)\) is calculated based on the continuity of the concentrations at the junction nodes, and it depends on the cell shape. For example, for a cell spanning 2 junctions, we denote node P as the reference point of polarity cue concentration with the concentration as \(c_0\) (Fig. \ref{fig1}B), and then the concentration profiles for all the arms and the middle segment can be written as

\begin{equation} \label{c_i_2junc}
\begin{split}
&c_{1/2}(x_{1/2}) = c_0 \, \exp{\left(-\frac{u_{1/2} \, x_{1/2}}{D}\right)} \\
&c_m(x_m) = c_0 \, \exp{\left(-\frac{u_m \, x_m}{D}\right)} \\
&c_{3/4}(x_{3/4}) = c_0 \, \exp{\left(\frac{-u_m \, d}{D}\right)} \, \exp{\left(-\frac{u_{3/4} \, x_{3/4}}{D}\right)} 
\end{split}
\end{equation}

The total concentration of the polarity cue is conserved within the cell
\begin{equation} \label{c_tot_2junc}
c_{tot}=\int_0^d c_m(x_m) \ dx_m + \sum_{j=1}^4 \int_0^{x_j} c_j(x_j) \ dx_j
\end{equation}
where the sum over \(\{j\}\) represents the contributions of the free arms. By substituting Eq.\ref{c_i_2junc} into Eq.\ref{c_tot_2junc}, we can obtain \(c_0\) as

\begin{eqnarray}
c_0 = \frac{c_{tot}}{D} \Large\left[ \frac{1-\exp{\left(-\frac{u_m\,d}{D}\right)}}{u_m} + \frac{1-\exp{\left(-\frac{u_1\,x_1}{D}\right)}}{u_1} + \frac{1-\exp{\left(-\frac{u_2\,x_2}{D}\right)}}{u_2} \right.\\ \nonumber
\left.+ \exp{\left(-\frac{u_m\,d}{D}\right)} \, \left(\frac{1-\exp{\left(-\frac{u_3\,x_3}{D}\right)}}{u_3} + \frac{1-\exp{\left(-\frac{u_4\,x_4}{D}\right)}}{u_4}\right) \Large\right]^{-1}
\end{eqnarray}

Generally, one can arbitrarily choose the reference point of the concentration field, denoted by \(c_0\), at one of the junctions. Then the concentration in segment \(i\) (\(i\) could be either the free arm or the node connection segment) can be written as

\begin{equation} \label{c_i}
c_i(x) = c_0 \, \exp{\left(-\frac{\sum_{k} u_k \, d}{D}\right)} \, \exp{\left(-\frac{u_i x}{D}\right)}
\end{equation}

\noindent
where \(\{k\}\) denotes the node connection segments between the segment \(i\) and the reference point. The direction of \(u_k\) is towards the reference point.

The total amount of the polarity cue is conserved within the cell
\begin{equation} \label{c_tot}
c_{tot} = \sum_{l} \int_0^d c_l(x_l) \ dx_l + \sum_{j} \int_0^{x_j} c_j(x_j) \ dx_j
\end{equation}
where \(\{l\}\) represent the nodes connected by full segments (of length $d$) and \(\{j\}\) represent the free arms.
From Eqs.\ref{c_i},\ref{c_tot} the value of $c_0$ can be found
\begin{equation} \label{c_0}
c_0 = \frac{c_{tot}}{D} [ \sum_l exp(-\frac{\sum_{k,l}u_{k,l}\,d}{D}) \, \frac{1-exp(-\frac{u_l\,d}{D})}{u_l} + \sum_j exp(-\frac{\sum_{k,j}u_{k,j}\,d}{D}) \, \frac{1-exp(-\frac{u_j\,x_j}{D})}{u_j} ]^{-1}
\end{equation}

where the summation \(k, l/j\) is for the nodes connected by segments \(\{k\}\) that are between the segment \(l/j\) and the reference point.

By substituting the expression of the inhibitor concentration (Eq.\ref{c_i}) into the expression of the steady-state local actin flows (Eq.\ref{v_ss}), we can obtain an implicit equation group of \(v_i\) \((i=1,2,...,N)\)

\begin{equation} \label{v_i_group}
v_i = \beta \frac{1}{ 1 + c_0 \, exp(-\frac{\sum_{k} u_k \, d}{D}) \, exp(-\frac{u_i x}{D}) }
\end{equation}

\noindent
where \(u_i\), \(u_k\) and \(c_0\) is given by Eq.\ref{u_i}, Eq.\ref{u_k} and Eq.\ref{c_0}, respectively. For the 2-junction case, the equation group of \(v_i\) \((i=1,2,3,4)\) is obtained by substituting Eq.\ref{c_i_2junc} into Eq.\ref{v_ss} (see Eq.S-1 in Supplementary Information section 1 as an example).

By numerically solving Eq.\ref{v_i_group}, we can determine the values of \(v_i\) corresponding to a specific set of \(\{x_i\}\). Following the initialization of \(x_i\), \(n_i\) and \(v_i\) of each arm, their temporal evolution during the migration process can be obtained through numerical integration of Eq.\ref{x_i}, Eq.\ref{n_i} and Eq.\ref{v_i}. In this study, we employed the symmetric initial condition.

\section*{Results and analysis}

\subsection*{Calculated shape dynamics of migrating cells}

\begin{figure}[t!]
    \centering
    \includegraphics[width=1\textwidth]{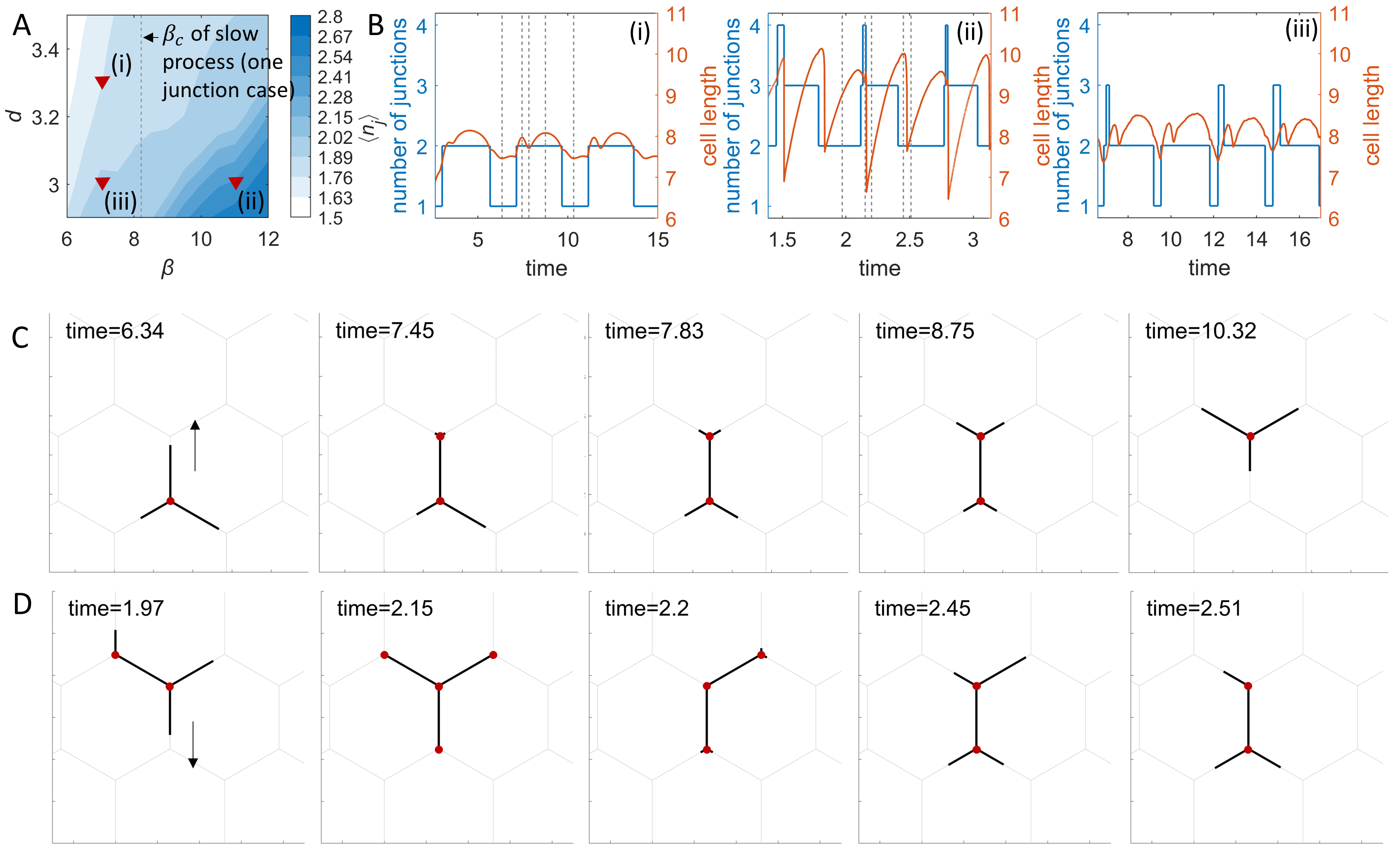}
    \caption{Analysis of the number of junctions in the migration process. A) The \(\beta-d\) phase diagram of the time average of the number of junctions. Gray dashed line represents the critical \(\beta\) of the slow process in the 1-junction case. B) The dynamics of the number of junctions and total cell length for the sections (i-iii) in (A). (i) \((\beta,d)=(6.5,3.3)\). (ii) \((\beta,d)=(11.0,3.0)\). (iii) \((\beta,d)=(7.0,3.0)\). C) Snapshots for the time stamps (gray dashed lines) in B(i). D) Snapshots for the time stamps (gray dashed lines) in B(ii). Black arrow points to the direction that the cell is moving forward to. Parameters: \(c=3.85, D=3.85, k=0.8, f_s=5, r=5, \kappa=20, \delta=250, \sigma=0.5\).}
    \label{fig2}
\end{figure}

We start by analysing the shape dynamics of our model cell moving on a hexagonal network, as function of the actin polymerization speed parameter \(\beta\) and the grid size \(d\). In  Fig. \ref{fig2}A we plot the time average of the number of junctions that the cell spans during its migration. Generally, cells have more arms when \(\beta\) is larger, since the protrusive forces that elongate the cell are stronger and a longer cell spans more junctions. Similarly, cells span more junctions and have more arms when \(d\) is smaller. 

We show in Fig. \ref{fig2}B the typical dynamics of the number of arms and the total cell length for the parameters indicated by (i) and (ii) (red inverted triangles) in Fig. \ref{fig2}A. We see that the larger number of junctions spanned by the cell is due to the larger mean values and the larger fluctuations in the total cell length. In Figs. \ref{fig2}C,\ref{fig2}D, we show snapshots for the time stamps (gray dashed lines) indicated in Figs. \ref{fig2}B(i),\ref{fig2}B(ii) respectively.

In Fig. \ref{fig2}A we indicate the critical value of $\beta$ above which we found that the cell tends to form the slow process (long-lived and very long protrusions) on a single junction \cite{ron2024emergent} (Gray dashed line), but the overall dynamics on the multiple junctions seems to be continuous as function of $\beta$, and we do not find any sharp change in the behavior. The critical \(\beta\)s for the cell to polarize on a 1D line and on different number of junctions are lower than the range that we used in this study. Their derivation and their values for the range of \(d\) that we used in this study are given in the SI section S-2 and Fig. S-1. We find that the critical beta, above which the cell polarizes and leaves the junction, increases with the number of junctions that the cell spans and therefore its number of protrusions, due to the increased competition between the more numerous leading edges, which makes it more difficult for the cell to polarize. This means that cells which migrate freely along linear tracks may get trapped, and become non-persistent (there is a residual diffusive migration), when moving inside the network and spanning junctions (thereby possessing multiple arms).

\subsection*{Theoretical large-scale migration characteristics of cells}

\begin{figure}[b!]
    \centering
    \includegraphics[width=\textwidth]{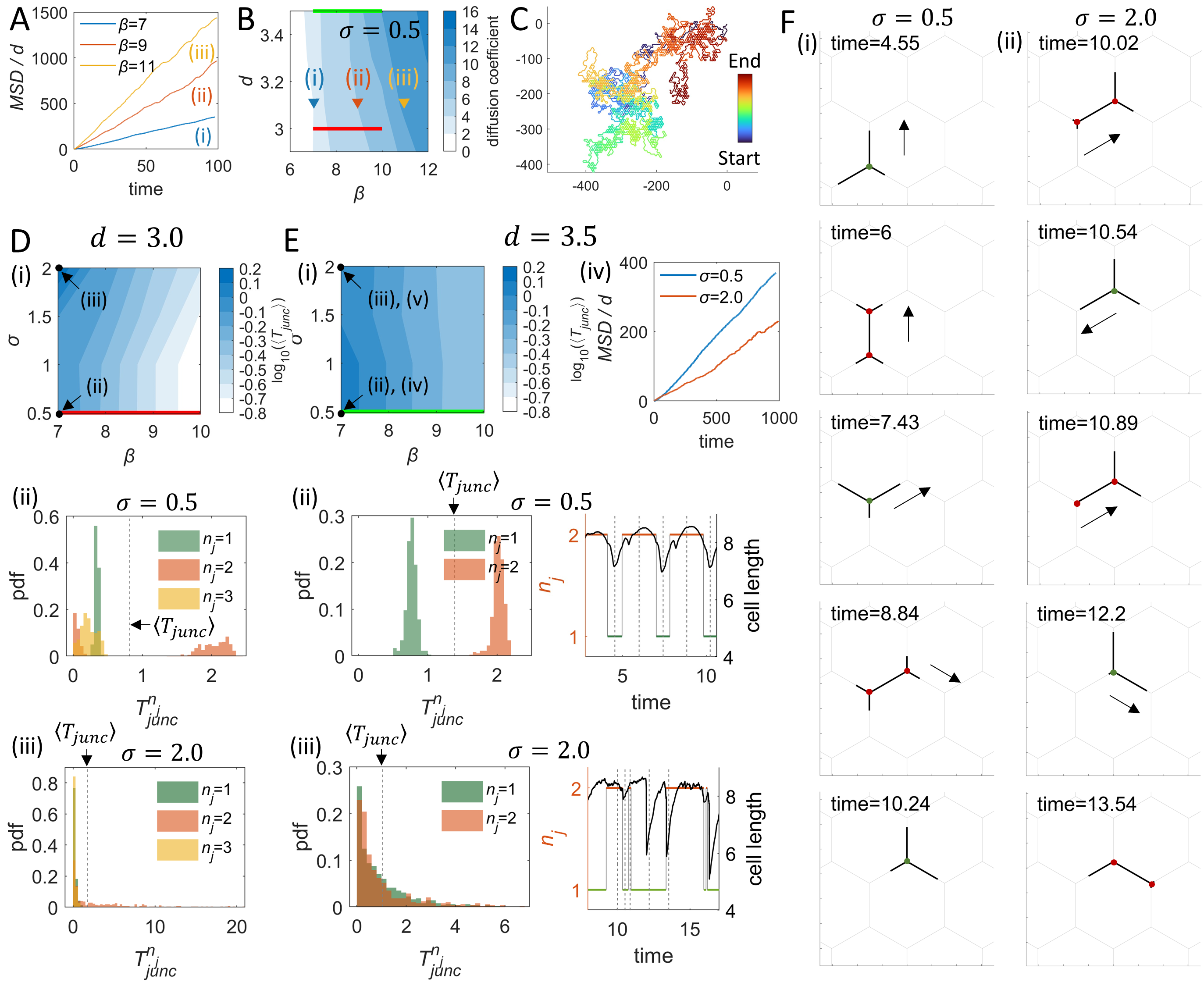}
    \caption{Analysis of the \(MSD\) of the cell's centroid and junction residency time. (A) Time series of \(MSD\) for \(\beta=7, 9, 11\) and \(d=3.1\). (B) The \(\beta-d\) phase diagram of the diffusion coefficient of the cell’s centroid ($D_c$, Eq.\ref{msd}). Red/Green line: the equivalent ranges of \((\beta, d)\) values spanned by Red/Green line in D(i)/E(i). Maximal simulation time: \(T=100\). (C) A typical trajectory of cell’s centroid during a long run for \(\beta=9\) and \(d=3.1\). Maximal simulation time: \(T=10000\). Other parameters: \(c=3.85, D=3.85, k=0.8, f_s=5, r=5, \kappa=20, \delta=250, \sigma=0.5\). (D) (i) The \(\beta-\sigma\) phase diagram of \(\langle T_{junc} \rangle\) (in log scale) for \(d=3.0\). (ii) Distribution (probability density function, pdf) of \(T_{junc}^{n_j}\) for \(\beta=7\), \(d=3.0\), \(\sigma=0.5\). (iii) Distribution of \(T_{junc}^{n_j}\) for \(\beta=7\), \(d=3.0\), \(\sigma=2.0\). (E) (i) The \(\beta-\sigma\) phase diagram of \(\langle T_{junc} \rangle\) (in log scale) for \(d=3.5\). (ii) Left: Distribution of \(T_{junc}^{n_j}\) for \(\beta=7\), \(d=3.5\), \(\sigma=0.5\). Right: Time series of the number of junctions and total cell length for \(\beta=7\), \(d=3.5\), \(\sigma=0.5\). (iii) Left: Distribution of \(T_{junc}^{n_j}\) for \(\beta=7\), \(d=3.5\), \(\sigma=2.0\). Right: Time series of the number of junctions and total cell length for \(\beta=7\), \(d=3.5\), \(\sigma=0.5\). (iv) Time series of \(MSD\) for two different levels of noise, with \(\beta=7\) and \(d=3.5\). (F) Simulation Snapshots. (i) Snapshots for the time stamps (gray dashed lines) in the right panel of E(ii). (ii) Snapshots for the time stamps (gray dashed lines) in the right panel of E(iii). }
    \label{fig3}
\end{figure}

Next, we investigate the large scale migration characteristics of the cells on the hexagonal network. We calculated the mean-squared-displacement (MSD) of the center-of-mass (see SI section S-3, Eq.S-14) of the cells during migration. We plot in Fig. \ref{fig3}A the time series of \(MSD\) for different values of \((\beta)\) and fit them by the power-law function \(MSD=Kt^\alpha\). We find that \(\alpha \approx 1\) for all the sets, corresponding to regular diffusive motion. For this two-dimensional Brownian motion, the \(MSD\) is
\begin{equation}
MSD(t) = 4D_ct
\label{msd}
\end{equation}
where \(D_c\) is the diffusive coefficient of the cell calculated by using the linear fit to the time series of the \(MSD\). In Fig. \ref{fig3}B we plot this diffusion coefficient as the function of \(\beta\) and \(d\) in Fig. \ref{fig3}B. We find that the diffusion coefficient increases as \(\beta\) and \(d\) increase. An example of typical long trajectory of the cell’s motion is shown in Fig. \ref{fig3}C.

In our study of cell migration over a single junction \cite{ron2024emergent} we found that the rate at which cells migrate over a junction generally decreases with increasing actin polymerization speed \(\beta\). This suggests that the trend shown in Fig. \ref{fig3}B can be explained by studying the rate at which the cell is able to migrate over a junction when spanning more than one junction. 

We start by plotting the average junction residency time of the cell, \(T_{junc}\), in Figs. \ref{fig3}D(i),E(i). The junction residency time is defined as the time it takes a cell to leave a junction since it was first occupied. We show the average junction residency time as function of \(\beta\) and \(\sigma\) (cellular internal noise, Eq.\ref{v_i}), for \(d=3.0\) (Fig. \ref{fig3}D(i)) and \(d=3.5\) (Fig. \ref{fig3}E(i)). The equivalent ranges of values spanned by Figs. \ref{fig3}D(i),E(i) are denoted by the red lines in Figs.\ref{fig3}B,\ref{fig3}D(i), and the green lines in Figs. \ref{fig3}B,\ref{fig3}E(i). In both cases, \(\langle T_{junc} \rangle\) decreases as \(\beta\) increases. This trend is consistent with the dependence of the escape time on \(\beta\) and \(\sigma\) in the one-junction case in our previous study \cite{ron2024emergent}: increase in \(\beta\), the polymerization activity at the arms' edges, increases the cell’s migration speed and decreases the residency time on a junction.  

For \(d=3.0\), the \(\langle T_{junc} \rangle\) increases as \(\sigma\) increases (Fig. \ref{fig3}D(i)). This is in agreement with the effect of noise for the migration of a cell over a single junction \cite{ron2024emergent}. The noise increases the residency time as it acts to decrease strength of the flows that polarize the cell. We show the average residency time of the cell over a specific number of junctions, \(\langle T_{junc}^{n_j} \rangle\) (where \(n_j\) represents the number of junctions that the cell is spanning), in Fig. S-2A. The distributions of \(T_{junc}^{n_j}\) are shown in Fig. \ref{fig3}D(ii),(iii) for low and high noise respectively. We see that the cells spanning 2 junctions are especially stable at low noise (Fig. \ref{fig3}D(ii)). The peak with smaller \(T_{junc}^{n_j=2}\) is due to the short time interval when \(n_j\) transits from 2 to 3, and the peak with larger \(T_{junc}^{n_j=2}\) is due to the long time interval when \(n_j\) transits from 2 to 1. This dynamics is shown in Fig. \ref{fig2}B(iii). At high noise, the average residency time increases due to the further increase in the residency time of these 2-junction spanning cells (Fig. \ref{fig3}D(iii)). At high noise, the residency time distribution becomes exponential, with a tail that reaches large values. 

When the grid size is larger, \(d=3.5\), we find that moderate noise can decrease the average junction residency time (Fig. \ref{fig3}E(i)). This trend is driven by the dynamics of the 2-junction spanning cells (SI section S-4). Due to the larger grid size, the cells form shorter arms that extend at the ends of the spanned junctions, and therefore noise can drive these short protrusive arms to loose their "grip" on one of the junctions. These dynamics and simulation snapshots are shown in Fig. \ref{fig3}E(ii,iii) and Fig. \ref{fig3}F, for the low and high noise cases respectively. 

This behavior is unique to cells spanning multiple junctions, whereby an increase in cellular noise can lead to a decrease in the time that a cell remain trapped in the junctions, and overall to a more frequent hopping between junctions. However, the reduction of the junction residency time does not correspond to faster global cellular diffusion. In Fig. \ref{fig3}E(vi) we show that the MSD for high noise is significantly smaller than that for low noise. The shorter junction residency time is driven by short bouts of the cell spanning two junctions and then loosing its "grip" on one of these junctions. However, these dynamics do not correlate with faster overall migration, as shown by the lower \(MSD\), and is rather in the form of rapid oscillations around the same average location. At higher values of noise we expect that the average junction residency time will eventually increase due to the loss of cellular polarization.

An alternative way to quantify the migration over the network is the "hexagonal residency time" of the cells, \(T_{hex}\), which is defined as the time from when the cell’s centroid enters a hexagonal cell of the network to when it leaves it and enter a new one. This quantification of the cellular migration is shown in the SI section S-5.

\section*{Comparison of the model with experiments}

\begin{figure}[t!]
    \centering
    \includegraphics[width=1\textwidth]{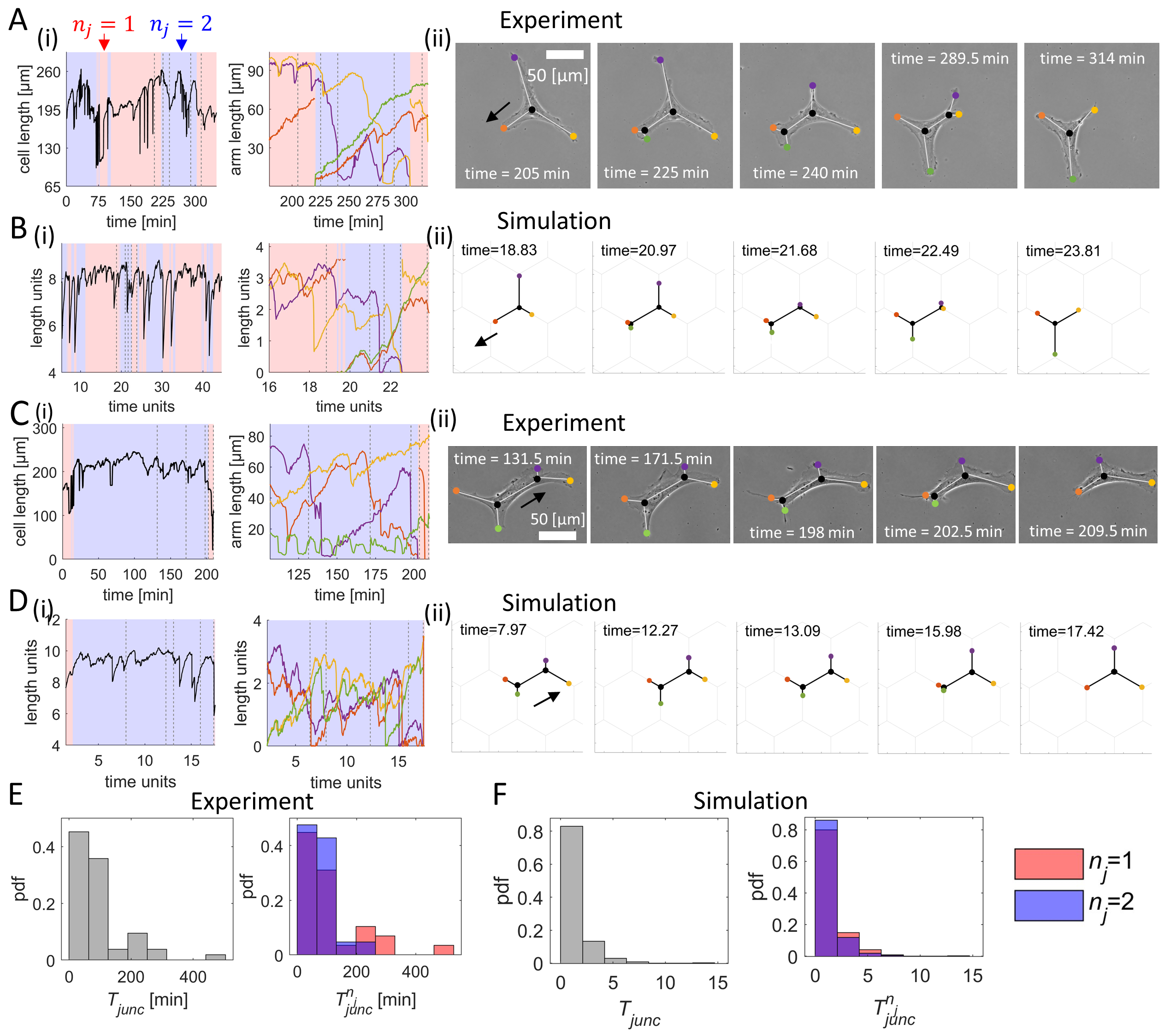}
    \caption{Comparisons between cell shapes and migration patterns in simulations and experiments on HUVEC. (A, C) Experiments on migrating Human Umbilical Vein Endothelial Cells (HUVEC). (B,D) Simulations of migrating  HUVEC. Hexagon side: 60~\textmu m. Inter-hexagon width: 20~\textmu m. (i) Dynamics of the total cell length and the arms' lengths. (ii) Snapshots of the cell, corresponding to gray dashed lines in (i). Key simulation parameters: \((\beta,d,\sigma,k)=(7.0,3.6,2.0,0.8)\) in (B), and \((\beta,d,\sigma,k)=(7.0,3.5,2.7,0.7)\) in (D). Other simulation parameters: \(c=3.85, D=3.85, f_s=5, r=5, \kappa=20, \delta=250\). (E) Distribution (probability density function, pdf) of \(T_{junc}\) (left) and distribution of \(T_{junc}^{n_j}\) (right) in the experiments, where the cells spans 1 or 2 junctions. A total of 9 cells across 5460.5 minutes were used to obtain these distributions. (F) Distribution of \(T_{junc}\) (left) and distribution of \(T_{junc}^{n_j}\) (right) from a long simulation with parameters of (B). }
    \label{fig4}
\end{figure}

\begin{figure}[t!]
    \centering
    \includegraphics[width=1\textwidth]{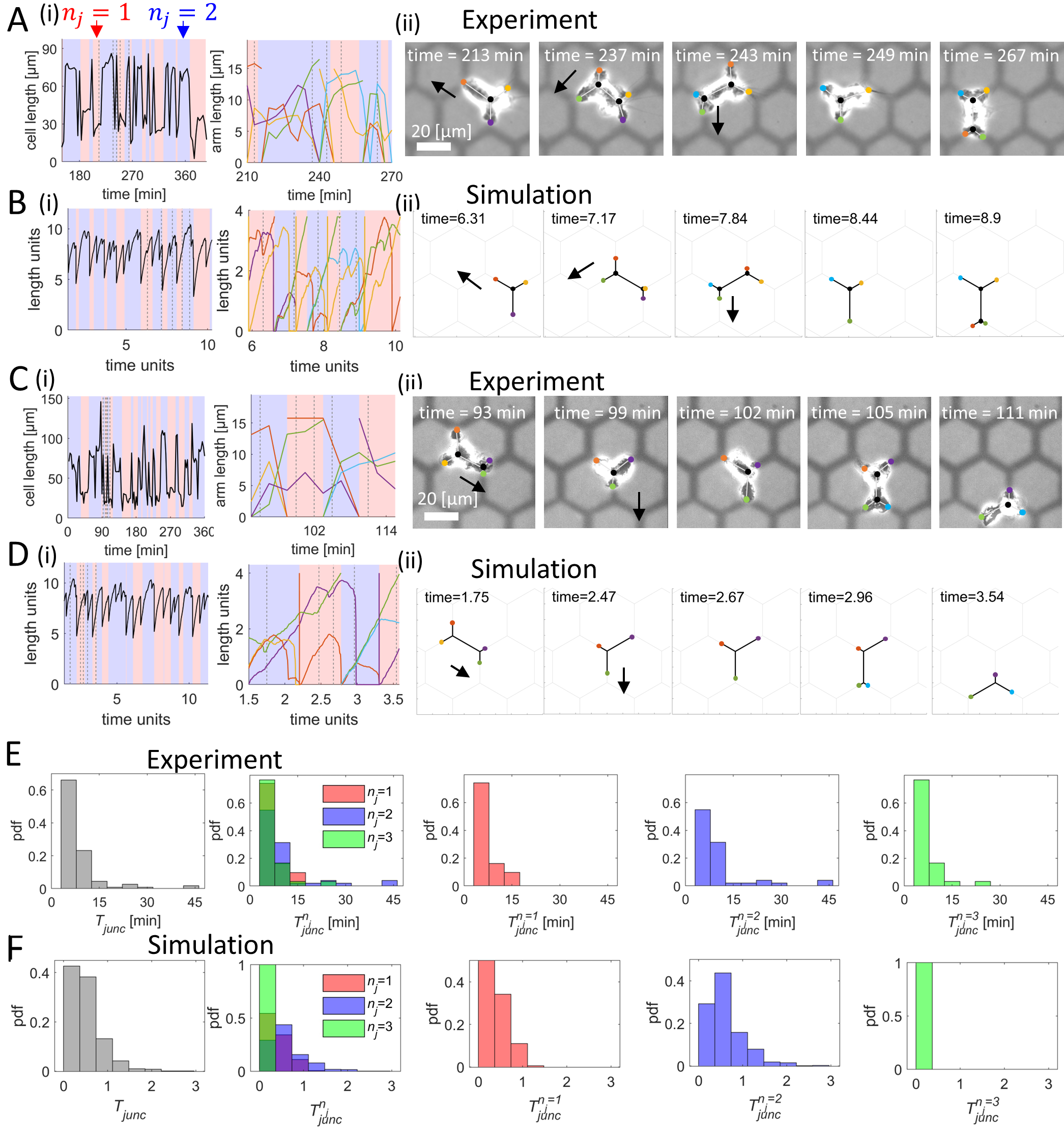}
    \caption{Comparisons between cell shapes and migration patterns in simulations and experiments on macrophage cells. (A, C) Experiments on migrating macrophages. Hexagon side: 12~\textmu m. Inter-hexagon width: 4~\textmu m. (B,D) Simulations of migrating macrophages. i) Dynamics of the total cell length and the arm lengths. ii) Snapshots of the cell, corresponding to gray dashed lines in (i). Key simulation parameters: \((\beta,d,\sigma)=(10.0,3.8,3.0)\) in B), and \((\beta,d,\sigma)=(10.0,4.0,2.5)\) in D). Other simulation parameters: \(c=3.85, D=3.85, k=0.8, f_s=5, r=8, \kappa=20, \delta=250\). (E) Distribution (probability density function, pdf) of \(T_{junc}\) (leftmost) and distribution of \(T_{junc}^{n_j}\) (right ones) in the experiments, where the cells spans 1 or 2 junctions. A total of 2 cells across 849 minutes were used to obtain these distributions. (F) Distribution of \(T_{junc}\) (leftmost) and distribution of \(T_{junc}^{n_j}\) (right ones) from a long simulation with parameters of (B). }
    \label{fig5}
\end{figure}

\begin{figure}[t!]
    \centering
    \includegraphics[width=1\textwidth]{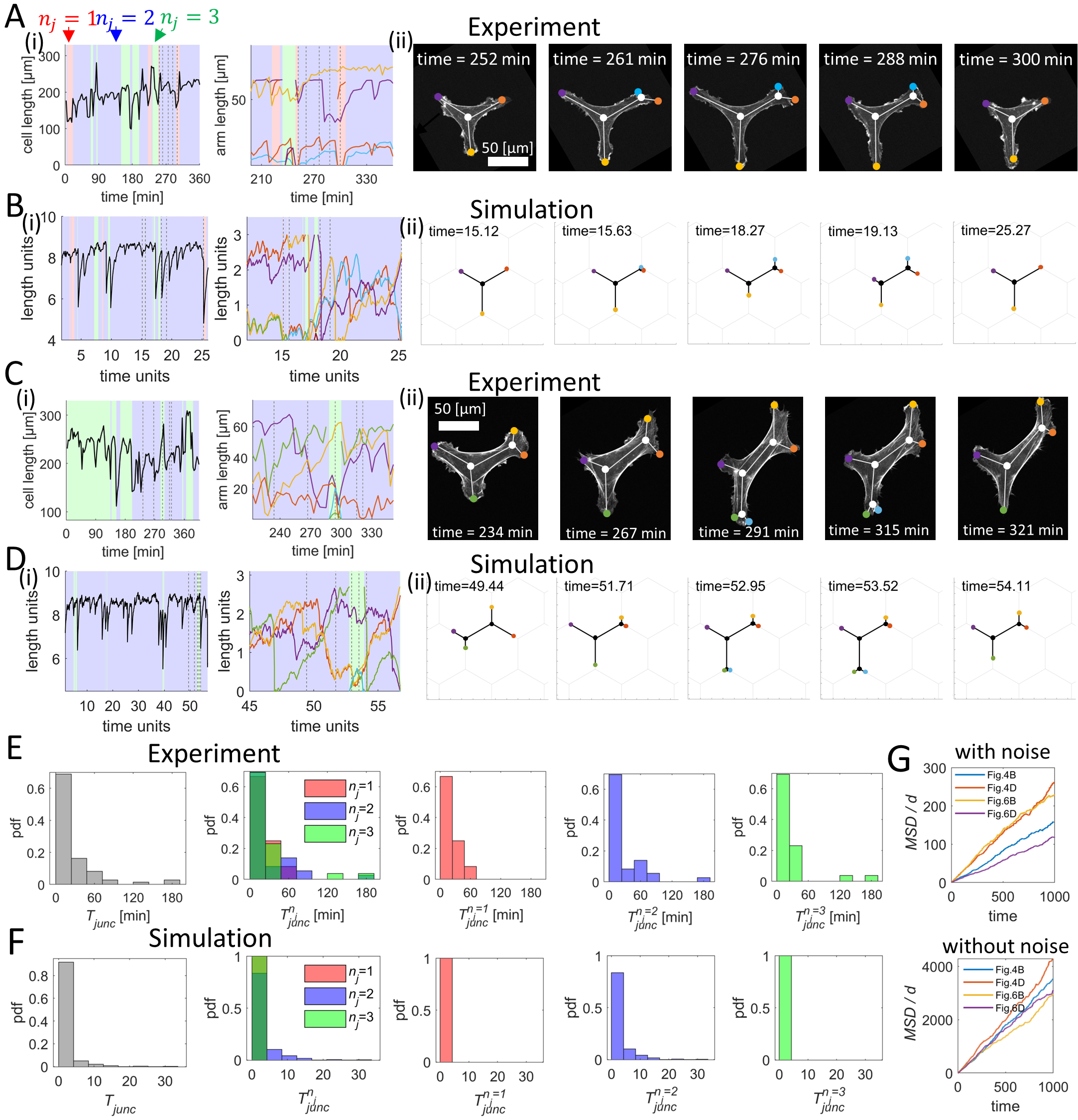}
    \caption{Comparisons between cell shapes dynamics in simulations and experiments for weakly motile cells (HUVEC). (A,C) Experiments on non-migrating HUVEC. (B,D) Simulations of non-migrating HUVEC. Hexagon side: 60~\textmu m. Inter-hexagon width: 20~\textmu m. i) Dynamics of the total cell length and the arm lengths. ii) Snapshots of the cell, corresponding to gray dashed lines in (i). Key simulation parameters: \((\beta,d,\sigma)=(7.0,3.0,1.8)\) in (B), and \((\beta,d,\sigma)=(7.0,2.7,2.1)\) in (D). Other simulation parameters: \(c=3.85, D=3.85, k=0.8, f_s=5, r=5, \kappa=20, \delta=250\). (E) Distribution (probability density function, pdf) of \(T_{junc}\) (leftmost) and distribution of \(T_{junc}^{n_j}\) (right ones) in the experimental regime in which the cells spans 1, 2 or 3 junctions. A total of 2 cells across 1287 minutes were used to obtain these distributions. (F) Distribution of \(T_{junc}\) (leftmost) and distribution of \(T_{junc}^{n_j}\) (right ones) from a long simulation with parameters of (B). (G) Comparisons of MSD of the simulated cells of Fig. 4 and Fig. 6, with the noise levels identical to those used in the corresponding simulations (top) or without the noise (bottom). }
    \label{fig6}
\end{figure}

We compare our model simulation results with experiments using Human Umbilical Vein Endothelial Cells (HUVEC) and macrophages cells placed on hexagonal networks of adhesive stripes, as the migration of both cell types is highly dependent on adhesion (Figs. \ref{fig4}-\ref{fig6}). We compare the calculated dynamics of the cell shape (number of junctions, cell length) as well as the distributions of junction residency times with the experimental observations. To extract these quantities from the experiments we used a convolutional neural network image analysis to compute the positions of the tips of cellular arms that emanate from the junctions spanned by the cells.

In Fig. \ref{fig4} we show the comparison between the results of simulations and experiments of migrating HUVEC. In these experiments, the size of the hexagonal network is such that the number of junctions spanned by the cells oscillates between 1 and 2. We present typical snapshots that illustrate the migration of cells on the hexagonal networks, and show that the overall dynamics of the cell length, individual arm lengths, and the number of spanned junctions is similar in experiments and simulations. We present two examples of cellular dynamics: in Fig. \ref{fig4}A,B the cells exhibit frequent fluctuations between spanning 1 and 2 junctions, while in Fig. \ref{fig4}C,D the cells are almost always spanning 2 junctions. Since the dimensions of the network are the same in both experiments of Fig. \ref{fig4}A,C (variation in the micropattern are negligible, of order $\sim1-5\%$), we may attribute the differences in cell migration pattern to variations in some cellular parameter that affects the cell length. In the simulations, we demonstrate in Fig. \ref{fig4}B,D that changes in the cell elasticity parameter $k$ affect the cell's total length, thereby changing its shape and the number of junctions it spans on the hexagonal network. A similar change in dynamics can also arise if the hexagonal network has slightly different size $d$ (see simulation results in SI section S-4, Fig. S-2).

The junction residency time distributions in both experiments and simulations have a broad, exponential form (Fig. \ref{fig4}E,F). Comparing to the distributions from the simulations at different noise levels (Fig. \ref{fig3}D(ii,iii),E(ii,iii)) this suggests that these cells are in a regime of significantly large internal noise. Overall, Fig. \ref{fig4} shows good qualitative agreement between the model and the experiments.

In Fig. \ref{fig5} (similar to Fig. \ref{fig4}) we show the comparison between the results of simulations and experiments of migrating macrophage cells. In these experiments, the size of the hexagonal network is such that the number of junctions spanned by the cells oscillates over the range of 1 to 3. The junction residency time distributions in the experiments and corresponding simulations are more tightly peaked compared with the HUVEC experiments, and have a much shorter duration (compare Figs. \ref{fig4}E,F and \ref{fig5}E,F). This suggests that these cells are in a regime of higher overall actin protrusive activity (our parameter $\beta$) and lower internal noise. These properties therefore correspond to macrophages being much more strongly polarized compared to the HUVEC, which is expected for these highly motile immune cells \cite{liu2020piezo1,paterson2022macrophage}.

In Fig. \ref{fig6} we compare between the results of simulations and two experiments of cells (HUVEC) that are stuck at the junctions, and are not globally motile over the timescale of the experiment. In our simulations this behavior occurs when the internal noise is too large for cells to efficiently polarize and migrate away from the junction. In the first example (Fig. \ref{fig6}A,B), the number of junctions oscillates over the range 1 to 3. In the second example (Fig. \ref{fig6}C,D), the number of junctions oscillates between 2 and 3.

We provide the normalized \(MSD\) of the simulated migrating HUVECs and non-migrating HUVECs (Fig. \ref{fig6}G), corresponding to the simulations used in Fig. \ref{fig4} and Fig. \ref{fig6} respectively, compared to the same simulations without internal noise. We demonstrate that weakly motile cells can arise from a combination of low values of actin polymerization (and retrograde flow, $\beta$) and large internal noise. Such cells tend to have long residency times, due to getting trapped over multiple junctions, and therefore have a low migration rate (and diffusion coefficient $D_c$, Eq.\ref{msd}) over the network.

\section*{Discussion}
In this work, we presented a theoretical model, and the corresponding experimental proof of concept, for cell shape dynamics during branching induced by migration inside a hexagonal network. The model describes the self-polarization of the cell due to the feedback between the actin polymerization activity at the leading edges of the cellular protrusions, and the global redistribution of a polarity cue that acts to inhibit local actin polymerization and protrusive activity. Here, we extend our recent description of cellular directional decision-making over a single junction \cite{ron2024emergent}, to allow cells spanning multiple junctions and have a highly ramified shape with multiple competing arms. Similar cell shapes are observed during cellular migration within tissues \textit{in vivo}\cite{liu2020piezo1,paterson2022macrophage}.

We compare the cellular dynamics predicted by the theoretical model with the migratory behaviour of two cell types that undergo stick-slip migration: HUVEC and macrophages. These cells were allowed to migrate on a hexagonal network of adhesive lines with sizes adapted to each cell: by constraining the cells to simultaneously span several junctions we trigger a high level of branching. The model provides a good qualitative description of the observed cellular shape dynamics. 

Using our model we investigate the role of noise in the polymerization activity, and the results suggest that cells have very noisy internal actin retrograde flows, which sometimes leads to their trapping at the junctions of the network. The latter suggests that there is an optimal range for the number of protrusions during cell migration: On the one hand these protrusions contribute to pathfinding and direct cellular migration as they sense the microenvironment. On the other hand, if the number of protrusions is too high, this will hinder efficient migration. This trade-off between environmental sensitivity and efficient migration may determine the optimal levels of acto-myosin contractility and protrusive activity, for each cell type. If the cell is trapped due to a noisy actin retrograde flow, a possible solution employed by cells to restart motility is to undergo a "reset" of the actin cytoskeleton, such as the previously described calcium-mediated actin reset \cite{wales2016calcium}.

Regarding the speed of migration, our data shows that macrophages are faster in exploring the hexagonal array, in comparison to HUVEC. This could be directly related to the surveillance function of macrophages, which requires the efficient exploration of the microenvironment, while preserving their migratory capacity \cite{paterson2022macrophage}. In addition, we found that the variations in the total cell length were higher in macrophages than in HUVEC. These findings might have implications regarding cell volume control, which also impacts cell migration \cite{zhang2022polarized,stock2024ph,schwab2012role}. Fast migrating macrophages would then require equally faster adaptation to the microenvironment, and therefore might need rapid changes in the function of molecules involved in cell volume control, whereas slower changes are expected in HUVEC. Therefore, in macrophages we expect to observe a prominent role for different cell volume regulators, such as mechanosensitive channels, tranporters/exchangers, and pumps \cite{zhang2022polarized,stock2024ph,schwab2012role}, which might contribute to provide the shape plasticity necessary for the efficient migration of these cells. Future studies should be performed to evaluate this hypothesis.   

The model presented in this work provides us with the framework to explore in the future more complex cell migration processes and it can be used to explore migration inside 2D and 3D networks composed of linear tracks linked in different complex geometries. This model extends the current conceptual and predictive frameworks to study cell migration in different geometries \cite{ron2020one,bruckner2023learning,ron2024emergent}. In addition, the present model can allow to study how such geometries affect the ability of cells to perform chemotaxis and haptotaxis in response to external gradients \cite{sengupta2021principles}. This will be most relevant to cells migrating through dense tissues, while guided by external signals, such as immune \cite{georgantzoglou2022two}, cancer \cite{paul2017cancer} cells, and others.

\section*{Experimental Methods}

\subsection*{Cell culture}

\textbf{Macrophages.} Bone marrow derived macrophages (BMDMs) were differentiated from myeloid precursors to macrophages in the span of 7 days. Briefly, bone marrows were isolated from C57BL/6 mice as described previously \cite{saez2017atp}, and were cultured in Corning culture dishes with 100 [mm] diameter using RPMI 1640 Medium, GlutaMAX™ Supplement, HEPES (10\% FCS) (Gibco, ThermoFisher), supplemented with 1\% Penicillin-Streptomycin (PenStrep, ThermoFisher) and 100 [µg/mL] of Macrophage Colony-Stimulating Factor (Bio-Techne GmbH). The cells were incubated at 37°C and 5\% CO2 for 7 [days], adding 5 [mL] of RPMI 1640 Medium, GlutaMAX™ Supplement, HEPES (10\% FCS), supplemented with 1\% Penicillin-Streptomycin, and 100 [µg/mL] of Macrophage Colony-Stimulating Factor (M-CSF) on days 3 and 5. On day 7, cells were detached using ethylenediaminetetraacetic acid (EDTA) 5 [mM] (ThermoFisher), and subsequently plated in the micropatterns.

\textbf{HUVECs.} The culture of Human Umbilical Vein Endothelial Cells (HUVEC) was performed as described previously \cite{ron2024emergent}. Briefly, HUVEC were acquired from PromoCell (C-12203) were used between passages 5 and 12. These cells were originally isolated from the vein of the umbilical cord of pooled donors. The HUVEC were cultured in full endothelial growth medium that included the Basal Medium and the SupplementMix (PromoCell), supplemented with 1\% Penicillin-Streptomycin (PenStrep, ThermoFisher), on 100 [mm] diameter Corning culture dishes, coated with fibronectin from bovine plasma (1 [µL/mL]) for 20 [min] at 37°C and 5\% CO2. The dishes were washed once with Dulbecco's Phosphate-Buffered Saline (DPBS, 1X, Gibco, ThermoFisher) prior to the culture. Once cultured, the cells were incubated at 37°C and 5\% CO2 for 2 [days] until reaching a confluency of 80\%. Once the monolayer is formed, the cells were passaged by detaching the cells with 3 [mL] of TrypLE Express (with Phenol Red, 1X, Gibco, ThermoFisher) for 2 [min] and then plating 1,5x106cells/[mL] in a new coated Corning culture dishes with 100 [mm] diameter in a total of 10 [mL] new full endothelial growth medium. 

\subsection*{Micropatterning}
The micropatterning was performed using the photopatterning technique, possible with a Digital Micromirror Device (Primo™, ALVEOLE) coupled to a T2i Eclipse microscope (Nikon). A Polydimethylsiloxan (PDMS) stencil with a circle area of 5 [mm] was placed inside a bottom glass dish (FluoroDish, World Precision Instruments), which is a 35 [mm] diameter glass bottom dish, previously cleaned with a PlasmaCleaner (PDC-32G-2, Harrick Plasma) for 5 [min]. 
Afterwards, the area within the stencil was coated for 1 [hr] at room temperature with 1 [mg/mL] pLL-PEG, in the case of the HUVEC, or fluorescent PEG atto 633 (SuSoS AG), in the case of macrophages. For HUVEC and Macrophages, the remaining pLL-PEG was washed with sterile water. 
Then in both cases, 10 µL of photo-activator (PLPP™, ALVEOLE) was added, and then degraded using UV illumination at a power of 600 [mJ/mm2], with the shape of the desired pattern. 
The pattern consisted of a hexagonal array, with a side of 60 [µm] and a width between hexagons of 20 [µm] in the case of HUVEC; and a side of 12 [µm] and a width between hexagons of 4 [µm] in the case of macrophages. These patterns were created using the software Inkscape 1.0.2-2 (see SI section S-6, Fig. S-4). For HUVEC, the  photo-activator (PLPP™, ALVEOLE) was washed with 1X DPBS and then the dish was coated for 20 min at 37°C with fibronectin (1 [μL/mL]) mixed with fibrinogen (1:10) to visualize the patterns, and then rinsed with 1X DPBS. For macrophages, the remaining  photo-activator (PLPP™, ALVEOLE) was washed out with 1X DPBS and where left uncoated to let the cells adhere to the glass of the bottom of the dish. After this 1X DPBS was removed and replaced with the corresponding media. Finally, the cells were plated as it follows. 10.000 HUVEC were seeded and left overnight (around [16h]) at 37°C and 5\% CO2 to ensure their adhesion to the plate. In the case of macrophages, 10,000 cells were plated and imaged around 6-8 [hr] later.

\subsection*{Transfection}
HUVEC were cultured until reaching a confluence of 80\%, and detached as mentioned above in the section for cell culture. Then the cells were transfected using the 4D-NucleofectorTM X Unit (Lonza), with 1 [µg] of F-tractin GFP plasmid (Plasmid \#58473 Addgene) and the P5 Primary Cell 4D-NucleofectorTM X Kit to then electroporate corresponding to the transfection program (CA-167) to then keep overnight (around 16 [hr]) in culture previous to use.

\subsection*{Live-cell imaging}
The imaging for the phase contrast movies was performed using an inverted microscope (Leica Dmi8) equipped with an APO 10x/0.45 PH1, FL L 20x/0.40 CORR PH1, and APO 40x/0.95 objectives. Images were recorded with an ORCA-Flash4.0 Digital camera (Hamamatsu Photonics) using the MetaMorph Version 7.10.3.279 software (Molecular Device). The actin fluorescent videos were performed with a Nikon Eclipse TiE equipped with a 40x Plan Fluor Phase objective. Images were recorded with an Photometrics Prime 95B (back-illuminated sCMOS, 11 µm pixel-size, 1200x1200 pixels) connected to a Yokogawa CSU W-1 SoRa in dual-camera configuration connected to a Confocal disk with 50 µm pinholes using the VisiView v4 software. During every acquisition, cells were kept at 37°C and 5\% CO2. For HUVEC, the acquisition was done with the FL L 20x/0.40 CORR PH1, actin visualization was done with 40x Plan Fluor Phase objective of the Nikon Eclipse TiE, both where imaged with a time interval of 3 [min], the rest of the phase contrast movies where imaged with the APO 40x/0.95 objective with a time interval of 30 [sec]. For macrophages, the acquisition for movies was done with an APO 40x/0.95 objective, with a time interval of 3 [min].

\subsection*{Tracking and neuronal network}
The coordinates of the tips of the cells across the junctions were monitored as it follows. Neural networks, with the ResNet50 architecture presented in DeepLabCut (version 2.3.0) \cite{mathis2018deeplabcut,nath2019using}, were trained to analyze the movement of the cells and extract these coordinates. DeepLabcut was installed in an Anaconda environment (Conda version 22.9.0) following the instructions on the corresponding documentation (https://github.com/DeepLabCut/DeepLabCut). The single-animal option was the one used in the analysis, as there was only one cell per movie. The “config.yaml” file was modified to include the different bodyparts (i.e, tips of cells and junctions) and it was added the skeleton that stated the relationship between each tip and the junction where it was expected to be found, to help the performance of the prediction.
For labelling the movies, different frames were extracted from the movies in regular quantities of 12 units. This means that if the video was 120 frames long, the 120 would be divided by 12 and the frames would be extracted 10 by 10 accordingly. Labels were added by hand in both the tips of the cells and the junctions, in each of the extracted frames, using DeepLabCut’s GUI. 
For HUVEC, a unique neural network was developed for all the movies as the similarities between the movies facilitated the generalisation of the analysis, training a network on the labelled data for 175,000 iterations.
For macrophages, neural networks were developed per single movie as the assumptions made by the network didn’t generalise well between different movies. This was likely due to the fact that macrophages were more motile than HUVEC, which increased the differences between movies. Then, each neural network was trained during 300,000 iterations on the labelled data, which provided a reliable basis for each macrophage movie analysed.
The data provided by the neural networks consists of three columns per label (X position, Y position and the likelihood of the prediction) and a row per frame. This data was represented in a video using a cutoff for the likelihood of 0.6, supervising the predictions and correcting them by hand when the prediction wasn’t in the tip of the cell.
Also, the predictions for the position of the junctions, as they were a fixed point, were substituted by the exact X and Y position of each junction.
To obtain the junction residency time, the following steps were followed. Frame by frame, the number of junctions in which the cell was spanning was annotated. With this information, the frequency of the amount of frames in which the cell was spanning a certain quantity of junctions was extracted. Finally, these distributions of the junction residency times were compared with those obtained from the simulation data.

\clearpage
\bibliography{bibliography}

\end{document}


\newcommand{\be}{\begin{equation}}
\newcommand{\ee}{\end{equation}}
\newcommand{\bea}{\begin{eqnarray}}
\newcommand{\eea}{\end{eqnarray}}
\newcommand{\nn}{\nonumber}
\renewcommand{\theequation}{S-\arabic{equation}}
\renewcommand{\figurename}{Fig.}
\renewcommand{\thefigure}{S-\arabic{figure}}

\title{Supplementary information - Shape dynamics and migration of branched cells on complex networks}
\author{Jiayi Liu$^{1,2}$, Javier Boix-Campos$^{3}$, Jonathan E. Ron$^{1,4}$, Johan M. Kux$^{3}$, Nir S. Gov$^{1}$ and Pablo J. S\'{a}ez$^{3}$}
\affiliation{$^{1}$Department of Chemical and Biological Physics, Weizmann Institute of Science, Rehovot, Israel}
\affiliation{$^{2}$Department of Physics, Yale University, New Haven, CT, USA}
\affiliation{$^{3}$Cell Communication and Migration Laboratory, Institute of Biochemistry and Molecular Cell Biology, Center for Experimental Medicine, University Medical Center Hamburg-Eppendorf, Hamburg, Germany}
\affiliation{$^{4}$Department of Physics, Technion, Haifa, Israel}

\maketitle

\section*{S-1.  Steady-state actin treadmilling flows for the 2-junction case}

For the 2-junction case, the equation group of \(v_i\) \((i=1,2,3,4)\) are

\begin{equation}
\begin{split}
v_{1/2} = \beta ( 1 + \frac{c}{D} \frac{exp(-\frac{u_{1/2} x_{1/2}}{D})}{\frac{1-exp(-\frac{u_m\,d}{D})}{u_m} + \frac{1-exp(-\frac{u_1\,x_1}{D})}{u_1} + \frac{1-exp(-\frac{u_2\,x_2}{D})}{u_2} + exp(-\frac{u_m\,d}{D}) \, (\frac{1-exp(-\frac{u_3\,x_3}{D})}{u_3} + \frac{1-exp(-\frac{u_4\,x_4}{D})}{u_4})} )^{-1} \\
v_{3/4} = \beta ( 1 + \frac{c}{D} \frac{exp(-\frac{u_{3/4} x_{3/4}}{D}) exp(-\frac{u_m\,d}{D})}{\frac{1-exp(-\frac{u_m\,d}{D})}{u_m} + \frac{1-exp(-\frac{u_1\,x_1}{D})}{u_1} + \frac{1-exp(-\frac{u_2\,x_2}{D})}{u_2} + exp(-\frac{u_m\,d}{D}) \, (\frac{1-exp(-\frac{u_3\,x_3}{D})}{u_3} + \frac{1-exp(-\frac{u_4\,x_4}{D})}{u_4})} )^{-1}
\end{split}
\end{equation}

\noindent
where we have rescaled \( c = \frac{c_{tot}}{c_s}\), and the actin flow speeds \(u_i\) are given by Eq.7 and Eq.8 in the main text. \\

\section*{S-2.  Critical local actin polymerization activity for symmetry breaking}

As in our previous model for cells on one junction, for cells that are spanning symmetrically on more junctions, there are also critical values of \(\beta\) which decides whether the cell can break the symmetry and start migrating. The \(\beta_c\) is determined by two lengths of the cell: the critical length of polarization due to the force balance, \(L_p\), and the critical length of polarization due to the redistribution of the polarity cue, \(L_c\). \\

\(L_p\) is calculated by the balance between the protrusion force and the elasticity of the cell. When the cell length is shorter than \(L_p\), the cell has a tendency to elongate. \(L_p\) has the same expression for the cases of different number of junctions:

\begin{equation}
L_p = \frac{1}{2} (1-c) + \frac{\beta}{2k} + \sqrt{c + ( \frac{1-c}{2} + \frac{\beta}{2k} )^2}
\end{equation}

\begin{figure}[b]
    \centering
    \includegraphics[width=1\textwidth]{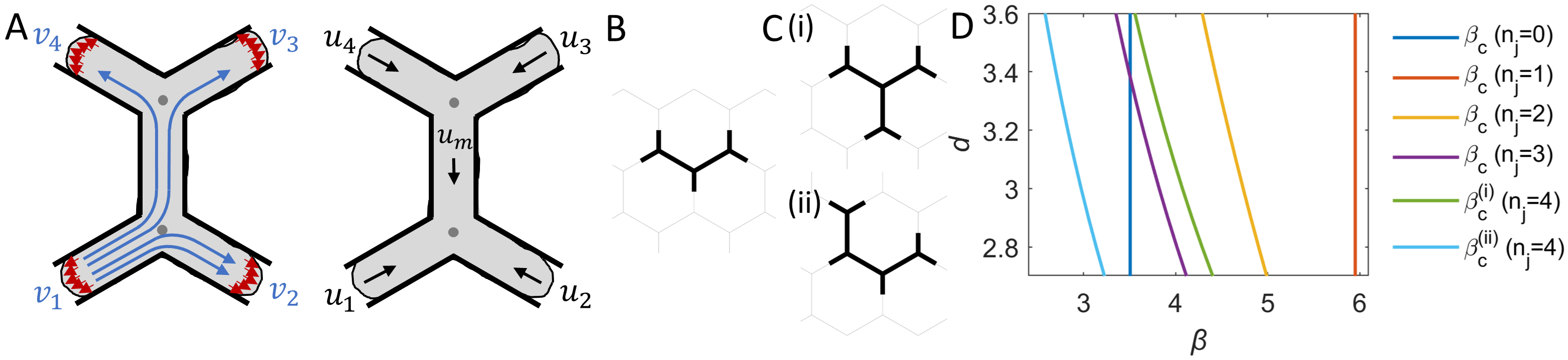}
    \caption{A) Left: Illustrations of the local actin treadmilling flows at each edge of the arms, and the split of the local actin flow of arm 1. Right: Illustrations of the net actin treadmilling flows in each segment of the cell. B) The cell shape for cells spanning 3 junctions. C) Possible cell shapes for cells spanning 4 junctions. D) Critical polymerization activity of cells spanning different number of junctions. Parameters: \(c=3.85, D=3.85, k=0.8\).}
    \label{figS1}
\end{figure}

On the other hand, when the cell length is longer than \(L_c\), the actin treadmilling flows of arms will deviate from the uniform solutions. In the instance of this symmetry breaking event, all the arms have the equal length, but there will be a small bias in some of the arms. One can derive \(L_c\) based on this event. Then the \(\beta_c\) is obtained by equating \(L_p\) and \(L_c\).

For cells migrating on a 1D straight line, and for cells on a single junction, the \(\beta_c\) is given by \cite{ron2020one,ron2024emergent}
\begin{equation}
\begin{split}
\beta_{c} (n_j=0) = \frac{D}{2c} + ck + \frac{ck^2}{2D} + \frac{ck-D}{2cD} \sqrt{D^2+2c(1+2c)Dk+c^2 k^2}
\end{split}
\end{equation}

\begin{equation}
\begin{split}
\beta_{c} (n_j=1) = \frac{D}{c} + ck + \frac{ck^2}{4D} + \frac{2D-ck}{4cD} \sqrt{4D^2+4c(1+2c)Dk+c^2 k^2}
\end{split}
\end{equation}

Next, we will derive the \(\beta_c\) in the 2-junction case as an example. We assume that \(x_i=l\), \(u_{1/2}=\epsilon\) and \(u_{3/4}=-\epsilon\) (Fig.\ref{figS1}A), then according to Eq.7 and Eq.8 in the main text, we have \(u_m=u_3+u_4=-2\epsilon\). By substituting these values to Eq.S-1, we obtain the actin treadmilling flows equations

\begin{equation}
\begin{split}
v_{1/2} = \beta ( 1 + \frac{c}{D} \frac{exp(-\frac{\epsilon l}{D})}{ \frac{1}{2\epsilon}[1-exp(-\frac{2\epsilon d}{D})] + \frac{2}{\epsilon}[1-exp(-\frac{\epsilon l}{D})] + exp(\frac{2\epsilon d}{D}) \frac{2}{\epsilon}[1-exp(\frac{\epsilon l}{D})]   } )^{-1} \\
v_{3/4} = \beta ( 1 + \frac{c}{D} \frac{exp(\frac{2\epsilon d}{D}) exp(\frac{\epsilon l}{D})}{ \frac{1}{2\epsilon}[1-exp(-\frac{2\epsilon d}{D})] + \frac{2}{\epsilon}[1-exp(-\frac{\epsilon l}{D})] + exp(\frac{2\epsilon d}{D}) \frac{2}{\epsilon}[1-exp(\frac{\epsilon l}{D})]   } )^{-1}
\end{split}
\end{equation}

Then we can write the global flow of arm 1, \(u_1=v_1-\frac{v_2}{2}-\frac{v_3}{4}-\frac{v_4}{4}\), as

\begin{equation}
\begin{split}
\epsilon = \frac{\beta}{2} ( 1 + \frac{c}{D} \frac{exp(-\frac{\epsilon l}{D})}{ \frac{1}{2\epsilon}[1-exp(-\frac{2\epsilon d}{D})] + \frac{2}{\epsilon}[1-exp(-\frac{\epsilon l}{D})] + exp(\frac{2\epsilon d}{D}) \frac{2}{\epsilon}[1-exp(\frac{\epsilon l}{D})]   } )^{-1} \\
- \frac{\beta}{2} ( 1 + \frac{c}{D} \frac{exp(\frac{2\epsilon d}{D}) exp(\frac{\epsilon l}{D})}{ \frac{1}{2\epsilon}[1-exp(-\frac{2\epsilon d}{D})] + \frac{2}{\epsilon}[1-exp(-\frac{\epsilon l}{D})] + exp(\frac{2\epsilon d}{D}) \frac{2}{\epsilon}[1-exp(\frac{\epsilon l}{D})]   } )^{-1} 
\end{split}
\end{equation}

Expand it in the first order in \(\epsilon\),

\begin{equation}
\epsilon = \frac{\beta c}{D} \frac{(d+l)(d+4l)}{(c+d+4l)^2} + O(\epsilon^2)
\end{equation}

Solve it for \(l\) and we can obtain the solution of the critical length for each arm,
\begin{equation}
l_c = \frac{ 8dD + c(8D-5d\beta+\sqrt{\beta(16cD-48dD+9d^2\beta)}}{8(\beta c -4 D)}
\end{equation}

The critical length of the cell is 
\begin{equation}
L_c = 4l_c+d = \frac{ 8dD + c(8D-5d\beta+\sqrt{\beta(16cD-48dD+9d^2\beta)}}{2(\beta c -4 D)} + d
\end{equation}

By equating \(L_p\) and \(L_c\), we can obtain the critical value of \(\beta\),
\begin{equation}
\begin{split}
\beta_{c} (n_j=2) = & \frac{1}{2c(4D+3dk)} [ 16D^2 + k^2 c(c-3d)(3d+1) + 4Dk(2c^2+3d-3cd) \\
            & + (4D-kc-3kd) \sqrt{16D^2 + 8c(2c-3d+1)Dk + c^2 (3d+1)^2 k^2} ]
\end{split}
\end{equation}

By similar derivations, we can obtain the critical \(\beta\) for the 3-junction case and for the 4-junction case (Fig.\ref{figS1}B-C),

\begin{equation}
\begin{split}
\beta_{c} (n_j=3) = & \frac{1}{24c(5D+6dk)} [ 400D^2 + 9k^2 c(c-8d)(8d+1) + 120Dk(c^2+4d-4cd) \\
            & + (20D-3kc-24kd) \sqrt{400D^2 + 120c(2c-8d+1)Dk + 9c^2 (8d+1)^2 k^2} ]
\end{split}
\end{equation}

\begin{equation}
\begin{split}
\beta_{c4}^{(i)} (n_j=4) = & \frac{1}{8c(8D+9dk)} [ 64D^2 + k^2 c(c-9d)(9d+1) + 8Dk(2c^2+9d-9cd) \\
                & + (8D-kc-9kd) \sqrt{64D^2 + 16c(2c-9d+1)Dk + c^2 (9d+1)^2 k^2} ]
\end{split}
\end{equation}

\begin{equation}
\begin{split}
\beta_{c4}^{(ii)} (n_j=4) = & \frac{1}{6c(4D+7dk)} [ 144D^2 + k^2 c(c-21d)(21d+1) + 12Dk(2c^2+21d-21cd) \\
                & + (12D-kc-21kd) \sqrt{144D^2 + 24c(2c-21d+1)Dk + c^2 (21d+1)^2 k^2} ]
\end{split}
\end{equation}

For the range of grid size \(d\) that we used in this study, we plot the \(\beta_c\) for different number of junctions in Fig.\ref{figS1}D. \\

\section*{S-3.  Mean-squared-displacement}
The mean-squared-displacement (MSD) of the center-of-mass of cells is given by

\begin{equation}
MSD(t) = \langle x(t)^2+y(t)^2 \rangle = \frac{1}{N_c} \sum_{i=1}^{N_c}(x_i(t)^2+y_i(t)^2)
\end{equation}

\noindent
where \(N_c\) is the number of cells that considered as an ensemble to be averaged (i.e., we run \(N_c\) repetitions of simulations), and \(x/y(t)\) is the \(x/y\) coordinate of the centroid of ith cell \((i=1,2,...,N_c)\) at time t in the simulation. Here we set \(N_c=1000\).\\

\section*{S-4.  Junction residency time for a specific number of junctions}
\begin{figure}[htbp]
    \centering
    \includegraphics[width=1\textwidth]{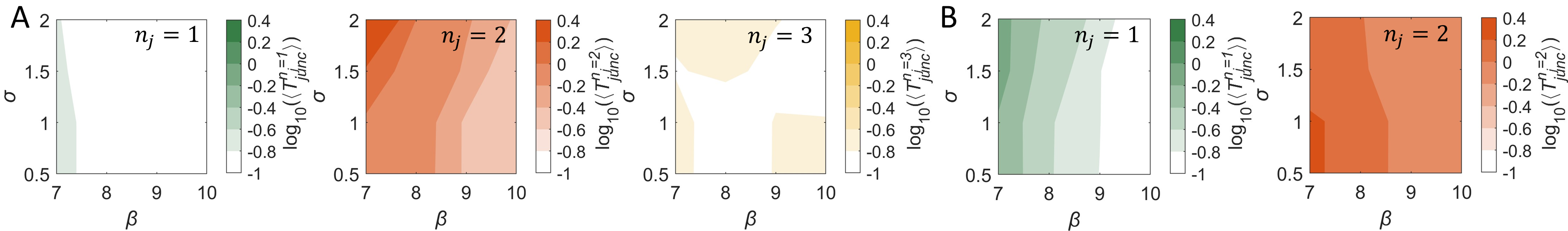}
    \caption{The \(\beta-\sigma\) phase diagrams of the junction residency time for a specific number of junctions (in log scale). A) \(d=3.0\). B) \(d=3.5\). Other parameters: \(c=3.85, D=3.85, k=0.8, f_s=5, r=5, \kappa=20, \delta=250\).}
    \label{figS2}
\end{figure}

We show the \(\beta-\sigma\) phase diagrams of the average junction residency time for \(d=3.0\) and \(d=3.5\) in the main text. The corresponding phase diagrams of the average junction residency time for a specific number of junctions are shown in Fig.\ref{figS2}A,B, respectively.

\section*{S-5.  Hexagonal residency time}

\begin{figure}[b]
    \centering
    \includegraphics[width=1\textwidth]{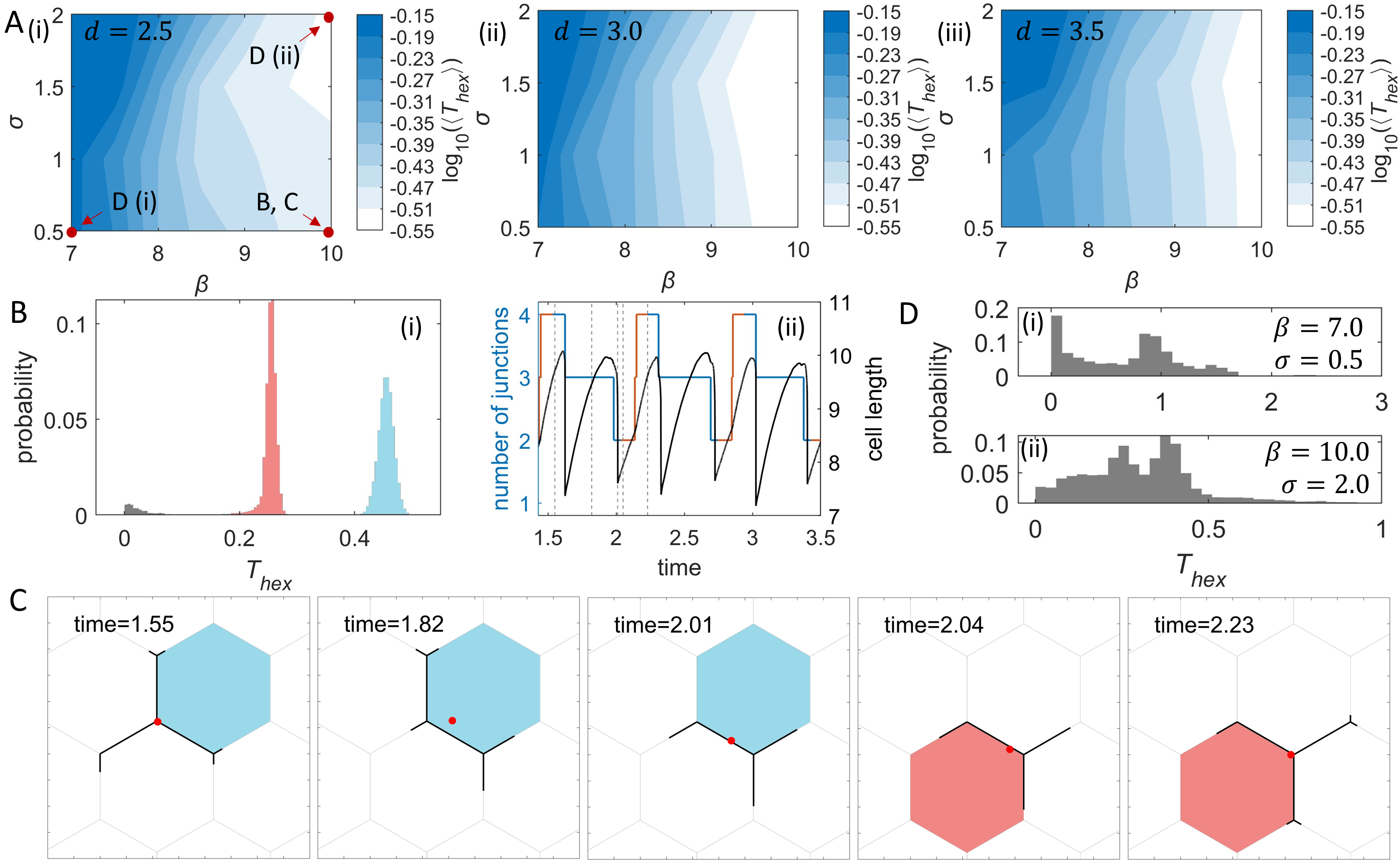}
    \caption{Analysis of the hexagonal residency time. A) The \(\beta-\sigma\) phase diagram of \(\langle T_{hex} \rangle\) (in log scale) for different sizes of grids. (i), (ii) and (iii) correspond to the grid size \(d=2.5\), \(d=3.0\) and \(d=3.5\). B) Analysis of the peaks of \(T_{hex}\) distribution for \((\beta,d,\sigma)=(10.0,2.5,0.5)\). i) The \(T_{hex}\) distribution. ii) Time series of the number of arms and cell length in a simulation. Blue/Red lines of the number of arms correspond to the Blue/Red peaks in (i). C) Snapshots of cell shape in a simulation, (i-v) corresponding to the five time stamps (gray dashed lines) in (ii) respectively. The red point indicates the centroid of the cell. The colored grid is the grid that the centroid is locating in, and Blue/Pink correspond to the Blue/Red peaks i of \(T_{hex}\) in B(i). D) The \(T_{hex}\) distribution for \(d=2.5\). (i) and (ii) correspond to \((\beta,\sigma)=(6.0,0.5)\) and \((\beta,\sigma)=(10.0,2.0)\) respectively. Parameters: \(c=3.85, D=3.85, k=0.8, f_s=5, r=5, \kappa=20, \delta=250\).}
    \label{figS3}
\end{figure}

The hexagonal residency time, \(T_{hex}\), is defined as the time from the cell’s centroid entering a hexagonal grid to leaving it and entering a new one. We show the phase diagram of the average hexagonal residency time, \(\langle T_{hex} \rangle\) (Fig.\ref{figS2}A). Consistent with the trend of \(\langle T_{junc} \rangle\) in the main text, the increase of \(\beta\) decrease \(\langle T_{hex} \rangle\). \\

We notice that for cells with large \(\beta\) and small \(\sigma\), the distribution of \(T_{hex}\) always have some peaks, which are correlated to some periodic patterns of the arm lengths and number of arms during the migration process. For example, the \(T_{hex}\) distribution for \((\beta,d,\sigma)=(10.0,2.5,0.5)\) has two discrete peaks (Blue/Red in Fig.\ref{figS3}B(i)), corresponding to the first and second half of the circulation of the arm lengths and number of arms (Blue/Red lines in Fig.\ref{figS3}B(ii)). In Fig.\ref{figS3}C, we show the snapshots for the shape and the centroid of the cell in a circulation. The centroid is marked by the red dot and the grid that it is occupying is colored in blue or pink, which corresponds to Blue/Red peak in Fig.\ref{figS3}B(i). Note that we ignore the small peak near \(T_{hex} \approx 0\) (Gray peak in Fig.\ref{figS3}B(ii)), which not indicates to periodic patterns. They originate from the small fluctuation of arm lengths when the cell shape is nearly symmetric and the centroid is close to the border of two adjacent grid. In this case, the slight drift of the centroid may lead to the stop of the residency on a grid. \\

The peaks of \(T_{hex}\) distributions become less significant as \(\beta\) decreases, because there are almost no oscillations in the cell length. An example for \((\beta,d,\sigma)=(6.0,2.5,0.5)\) is shown in Fig.\ref{figS3}D(i). The peaks shift left and connect with each other when \(\sigma\) is large, because the noise makes the centroid drifts more irregularly. An example for \((\beta,d,\sigma)=(10.0,2.5,2.0)\) is shown in Fig.\ref{figS3}D(ii).\\

\section*{S-6.  Micropatterns of HUVECs and macrophages}
\begin{figure}[htbp]
    \centering
    \includegraphics[width=.7\textwidth]{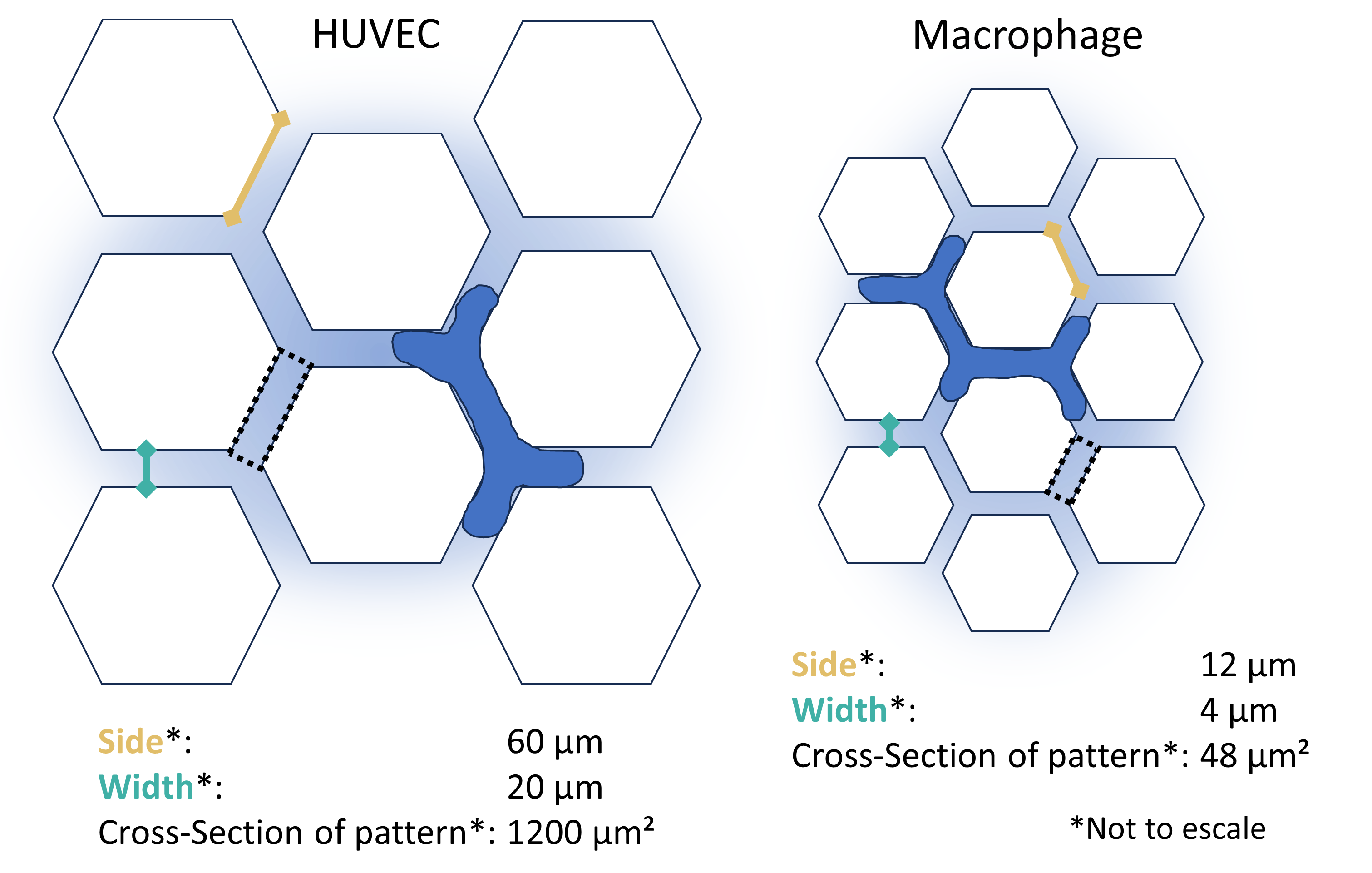}
    \caption{Micropatterns of HUVEC and macrophages. Left: Hexagonal micropattern of the HUVEC. Right: Hexagonal micropattern of the macrophages. Note the differences in the size of the hexagonal array for each cell type.}
    \label{figS4}
\end{figure}

The micropatterns of HUVECs and macrophages in our experiments has been performed as described the "Experimental Methods" section of the main text, and shown schematically in Fig.\ref{figS4}. Note that we have adapted the size of the hexagonal array to allow the HUVECs and macrophages to span simultaneously several junctions. Because macrophages are smaller than HUVECs we have reduced the size of the hexagons previously used \cite{ron2024emergent}, as described in Fig.\ref{figS4}.

\bibliography{bibliography_si}